\documentclass[12pt,preprint]{aastex}
\usepackage[percent]{overpic}
\usepackage{graphicx}
\usepackage{natbib}
\usepackage[dvips]{epsfig,color}
\usepackage{xr}
\usepackage{url}
\usepackage{subfigure}

\newcommand{\angstrom}{\mbox{\normalfont\AA}}
\def\kms{\ifmmode km\thinspace \mbox{s}^{-1}\else km\thinspace$\mbox{s}^{-1}$\fi}     
\def\deg{\ifmmode^\circ\else$^\circ$\fi}  
\def\arcs{\ifmmode {'' }\else $'' $\fi}  
\def\arcm{\ifmmode {' }\else $' $\fi}    

\def\lya{Ly-$\alpha$}

\def\h2{$\mbox{H}_2$}
\def\lapp{\ifmmode {_<\atop{^\sim}} \else {$_<\atop{^\sim}$}\fi} 
\def\gapp{\ifmmode {_>\atop{^\sim}} \else {$_>\atop{^\sim}$}\fi} 

\newcommand{\hb}{\hbox{H$\beta$}}

\begin{document}

\title{Recoiling supermassive black hole in changing-look AGN Mrk 1018}

\author{
D.-C. Kim\altaffilmark{1}, 
Ilsang Yoon\altaffilmark{1}, \&
A. S. Evans\altaffilmark{1,2}
}
\altaffiltext{1}{National Radio Astronomy Observatory, 520 Edgemont Road,
Charlottesville, VA 22903: dkim@nrao.edu, iyoon@nrao.edu}

\altaffiltext{2}{Department of Astronomy, University of Virginia, 530 McCormick Road, Charlottesville, VA 22904}

\textbf{Abstract}

\noindent
The spectral type of Mrk 1018 changed from Type 1.9 to 1 and returned back to 1.9 over a period of 40 years.
We have investigated physical mechanisms responsible for the spectral change in Mrk 1018
by analyzing archival spectral and imaging data.
Two kinematically distinct broad-line components, blueshifted and redshifted components, are found from spectral decomposition.
The velocity offset curve of the broad-line as a function of time shows a characteristic pattern.
An oscillating recoiled supermassive black hole (rSMBH) scenario is proposed to explain the observed velocity offset in broad emission lines.
A Bayesian Markov-Chain Monte Carlo simulation is performed to derive the best fit orbital parameters; 
we find that the rSMBH has a highly eccentric orbit with a period of $\sim$29 years.
The AGN activity traced by variation of broad H$\beta$ emission line is found to increase and decrease rapidly 
at the start and end of the cycle, and peaked twice at the start and near the end of the cycle. 
Extinction at the start and end of the cycle (when its spectral type is Type 1.9) is found to increase due to increased covering factor.
Perturbations of the accretion disk caused by pericentric passage 
can reasonably explain the AGN activity and spectral change in Mrk 1018.
Currently, the spectral type of Mrk 1018 is Type 1.9, and
we do not know if it will repeat a similar pattern of spectral change in the future,
but if it does, then spectral type will turn to Type 1 around mid 2020's.

\clearpage


\clearpage
\noindent
\section{Introduction}

\noindent
Supermassive black holes in the center of galaxies grow their black hole mass by accreting gas and dust in
the surrounding region, a process which results in them being observed as active galactic nuclei (AGN).
The AGNs are classified from Type 1 to intermediate types (Type 1.8 and 1.9) to 
Type 2 depending on the existence of broad emission lines in spectra.
Type 1 AGNs show both broad and narrow emission lines, while Type 2 AGNs only have narrow emission lines.
Type 1.8 and 1.9 AGNs show broad H$\alpha$ emission line, but lack broad H$\beta$ emission line (Osterbrock 1981).
Conventionally, these classifications can be explained by different viewing angles towards the accretion disk (Antonucci 1993): 
direct and tilted view onto the accretion disk yield Type 1 AGN and Type 1.8 to 2 AGN respectively.
Recently, a dozen AGNs have been reported to change their spectral types (so-called changing-look AGN) and
a few of them, including Mrk 1018 are known to have transitioned across the spectral type twice (Denney et al. 2014; MacLeod et al. 2016; McElroy et al. 2016).

Mrk 1018 is a galaxy merger as evidenced by tidal tails (Fig. 1).
The first spectroscopic observation was done by Markarian et al. (1977) on September, 1974 and
in the digitized spectra (Mickaelian et al. 2007), broad H$\alpha$ emission line is clearly seen, but broad H$\beta$ line is not visible.
Subsequent spectra were taken on November, 1976 (Afanas\'ev, Denisyuk, \& Lipovetskii, 1979), 
and the authors remarked that the presence of Seyfert properties in Mrk 1018 are uncertain and classified the spectra as Type 2.
Follow-up observations suggest the spectral type remained almost the same until the late 1970s (Afanas\'ev et al. 1980, 1982; Osterbrock, 1981).
The broad H$\beta$ emission line became visible and its spectral type changed to Type 1 around 1984 (Cohen et al. 1986)
and was observed to be Type 1.9 again in 2015.
Two mechanisms could explain the change of spectral type: 
i) obscuring material blocks the line of sight, or 
ii) change of AGN accretion activity. 
Observed flux variations in optical and X-ray data (McElroy et al. 2016; Husemann et al. 2016; LaMassa, Yaqoob, \& Kilgard 2017) led to suggestion that 
a decreased accretion rate in AGN was responsible for the changing-look in Mrk 1018.
In this paper, we investigate the physical mechanisms behind the change in the spectral type of Mrk 1018 by analyzing archival spectra and images.


\section {Spectroscopic Analysis}

\subsection {Spectral Decomposition}

The archival spectroscopic data (Fig. 2) we have collected span a 41 year period. 
The source of the data and epoch of the observations are listed in Table 1.
To trace variations in the broad H$\alpha$ emission line, spectra in Fig. 2 (except for Byurakan spectra) 
are normalized to have the same integrated flux value of the [O III]$\lambda 5007$\AA \ line 
under the assumption that the narrow [O III] emission line does not vary during the 41 year time span.
Among the spectra, only the SDSS and MUSE spectra are flux-calibrated.
The fiber diameter of the SDSS spectra is 3\arcsec \ and to match the SDSS fiber diameter,
the MUSE spectra were extracted within a 3\arcsec \ diameter aperture centered on the AGN position.
We find that the [O III]$\lambda 5007$\AA \ line flux of the MUSE spectra is about 30\% smaller than that of the SDSS.
Thus, we rescaled the MUSE spectra to match the SDSS spectra by multiplying the former by a factor of 1.3. 
After that, we compared the strength of Mg Ib and Na I D stellar absorption lines and found they have line strengths that agree to within $\pm$5\%.

The change in the accretion rate of the AGN causes changes in the continuum level that in turn causes changes in the strength, shape, and line-width of the broad-line.
Thus, an analysis of the broad emission line could shed some light on the variability of the AGN.
To extract physical information of the broad emission lines, spectral decompositions were performed for broad H$\alpha$ emission line using IRAF/Specfit package with 3 component fits:
i) power-law continuum, ii) broad emission line, and iii) narrow emission line.
We applied a Lorentzian profile for the broad H$\alpha$ emission line component that represents turbulent motion in the BLR and a Gaussian profile for the narrow H${\alpha}$, [N II], and [S II] line components.
The same Gaussian line width was used for the narrow emission lines of H${\alpha}$, [N II] and [S II], and
a fixed value of $1/3$ was used for the [NII]6548 to [NII]6583 line ratio.
For the broad H${\alpha}$ emission line, we first attempted a fit with a single Lorentzian, but it did not produce a good result
mainly because of blue or red asymmetric profiles in the broad-line.
Thus, we tried fits with multiple Lorentzians and found two Lorentzian components for the broad-line yielded the best result.

The result of spectral decomposition is plotted in Fig. 3
where black, magenta, blue, red, green, and orange lines represent data, model, broad-line blue component, broad-line red component, narrow lines, and absorption lines, respectively.
In the plot, we find two kinematically well separated broad-line velocity components: a blue component and red component.

\subsection {Interpretation of broad-line velocity offset curve}

The broad H$\alpha$ emission line velocity offsets for the blue and red components with respect to narrow H$\alpha$ line are listed in Table 1 and 
are plotted as a function of epoch in Fig. 5a, 
where blue, red, and black points represent broad-line velocity offsets for blue component, red component, and average of the blue and red components, respectively.
The velocity offset curve for the blue and red components show systematic change and suggests a bulk motion in the both components.

If the line of sight velocity differences between blue and red broad-line components are changing at each epoch,
it would be strong evidence that it is produced by rotating binary black holes (BBHs). 
However, the velocity difference between the blue and red broad-line components is about 2,000 \kms\ (Fig. 5a) and does not change significantly,
nor does it alternate in the manner observed in BBHs (Shen \& Loeb 2010).
Furthermore, if the BBHs are orbiting each other in a circular motion with respect to center of mass,
the larger black hole will rotate slower than the smaller black hole.
Black hole mass has a dependence of a square of broad-line width (Peterson \& Wandel 1999).
Thus, if they are orbiting each other, a broad-line component with larger velocity offset will have a smaller line width at each epoch, 
but this is not always true in the case of Mrk 1018 (see the 2nd and 3rd columns of Table 1).
The binary black hole scenario also has a difficulty to explaining the velocity offset curve at epoch 1984.08 and 2015.03 
that significantly deviates from sinusoidal motion.
A lack of pairs of narrow emission lines in the spectra is additional evidence disfavoring the BBH scenario.

Another scenario that could produce the asymmetric broad-line profile is a combination of inflow and outflow.
This scenario requires that both the inflow and outflow exist simultaneously at each epoch 
and that they periodically change their velocities: 
i.e. if the inflow velocity increases, then the outflow velocity has to decrease, and vice versa.
In addition, both the inflow and outflow velocities need to have sudden velocity jump in the negative direction (epoch $\sim$1985) or positive direction (epoch $\sim$2014).
This scenario was rejected since it is physically too unrealistic.

We propose that the characteristic velocity offset curve in Fig. 5a is originated by two kinematically distinct 
broad line regions (BLRs) in a recoiled supermassive black hole (rSMBH).
It is suggested that when two SMBHs coalesce at the final stage of a galaxy interaction,
{\it a merged SMBH} can recoil from the host galaxy due to anisotropic emission of gravitational waves (Peres 1962).
Recent simulations of merging black holes predict that the merged SMBH can attain a kick velocity of
a few hundred to a few thousands km s$^{-1}$ depending on mass ratios, spin magnitudes,
and spin orientations of the merging SMBHs (Campanelli et al. 2007; Schnittman 2007; Baker et al. 2008; Lousto \& Zlochower 2011; Blecha et al. 2016).
The rSMBH carries along with the broad-line region (BLR) and leaves the stellar nucleus behind;
it can be observable for tens of Myr as an offset AGN (Madau \& Quataert 2004; Loeb 2007; Blecha et al. 2011, 2016).
Recently, we have performed a systematic search for rSMBHs and find 26 rSMBH candidates (Kim et al. 2016) and one potential rSMBH (Kim et al. 2017).
Measured recoiling velocities for these candidates range from 80 \kms\ to 310 \kms\ and are significantly smaller than galaxy escape velocity (a few thousands \kms).
Thus, the rSMBHs will undergo oscillatory motion around the center of host galaxy until dynamical friction completely dampens the motion.
Given that Mrk1018 is a late-stage merger as shown in Fig. 1, the two SMBHs from the original galaxies may have already merged. 
The resulting merged SMBH may have recoiled and is presently undergoing oscillatory motion.

The BLR is thought to be tidally disrupted dusty clumps originating from the torus and bounded by the gravitational potential of a supermassive black hole (Wang et al. 2017).
It is hard to imagine that both kinematically distinct red and blue BLR components originate from the same torus.
We suspect that the two kinematically distinct BLRs we proposed are BLRs in the original galaxies and are gravitationally bound to the merged rSMBH.
In general, the orbiting BBHs have two spatially and kinematically distinct broad-line components and show periodic velocity shifts (i.e. Fig. 3 of Simi\'c \& Popovi\'c 2016).
The pattern of velocity shifts as a function of time is quite similar to the ones we have found in Fig. 3.
We do not know whether two BLRs moving along with the rSMBH are also merged to form a single BLR,
or still remain two kinematically distinct BLRs.
If the broad-line velocity offsets observed in Fig. 3 originated from the BLRs in the rSMBH, 
then it could be possible that they are two kinematically separated BLRs, not a merged single BLR.
%
%
%

\section{Modeling Broad-line Velocity Offset}
To test oscillating rSMBH hypothesis, we have modeled the broad-line velocity offset data using an eccentric orbit.
We placed the focus at the origin of orbital plane (($x,y)=(0,0)$) and pericenter and apocenter on the negative and positive sides on the $X$-axis.
We assume that at time $t_0$, the rSMBH passes the pericenter and orbits in clockwise direction with period $P$ and 
eccentricity $e=\sqrt{1-(b/a)^2}$ ($a$ and $b$ are the semi-major and semi-minor axis length, respectively). 
The line of sight is on the negative side of $X$-axis.
For elliptical orbits, the mean anomaly $M$ defined as $2\pi (t-t_0)/P$ is related with eccentric anomaly $E$ as follows:
\begin{equation}
\frac{2\pi (t-t_0)}{P} = E - e~\mbox{sin}(E).
\end{equation}
In such configuration, the $x$ and $y$ velocity components become 
\begin{eqnarray}
v_x & = &\frac{2\pi a}{P} \frac{\mbox{sin}(E)}{1-e~\mbox{cos}(E)}\\
v_y & = &\frac{2\pi b}{P} \frac{\mbox{cos}(E)}{1-e~\mbox{cos}(E)}.
\end{eqnarray} 
We have further assumed that the orbit spins around the orbital axis (i.e., $Z$-axis) in clockwise direction with period $Q$ and 
initial argument of periapsis $w_0$. 
For modeling, we used average velocity offset values of red and blue broad-line components
and allowed them to have an unknown constant velocity offset from the \textit{true} broad-line velocity. 
With this setup and inclination angle $i$, the line of sight velocity is given by 

\begin{equation}
v_{\mbox{\tiny los}} = \left(v_x \mbox{cos}(w) + v_y \mbox{sin}(w)\right)\times \mbox{sin}(i)  + v_{\mbox{\tiny off}}
\end{equation} where $w=2\pi (t-t_0)/Q+w_0$ and $v_{\mbox{\tiny off}}$ is the velocity offset from the \textit{true} 
broad-line velocity we have assumed. 

The inclination angle $i$ was found to be 49\deg\ based on modeling of $r$-band image with GALFIT (Peng et al. 2010)
and this value was fixed so that the model orbit is aligned with the galactic plane of Mrk 1018.
We decided to fix the inclination angle because the inclination $i$ is degenerated with semi-major axis parameter $a$ in 
Equations (2) - (4) such that only $a\times\mbox{sin}(i)$ can be constrained by the data. 
Thus, depending on the assumed value of the inclination angle, the semi-major axis length 
can be larger or smaller than the best fit value, 0.004 pc calulated from the fixed inclination angle of 49\deg.
In total, the model has seven free parameters: $P, t_0, e, a, Q, w_0, v_{\mbox{\tiny off}}$.

We adopt a Bayesian Markov chain Monte Carlo (MCMC) technique with $flat$ prior to find the best-fit parameters for the rSMBH orbit.  
The posterior probability is sampled using the Python package \texttt{emcee} (Foreman-Mackey et al. 2013).
In total, approximately 650,000 converged MCMC samples were used to characterized the posterior probability. The obtained MCMC samples 
has a reasonable acceptance rate (25\% on average) and small auto-correlation lag (5\% of the chain length on average), which indicates that
the samples are converged in practice.

The 1D and 2D marginal posterior probability distribution of model parameters are created with the Python package, \texttt{corner.py} (Foreman-Mackey 2016) and shown in Fig. 4.
Each diagonal plot shows 
1D marginal posterior distribution of the corresponding model parameter and off-diagonal plots show 2D joint marginal posterior 
distribution of two parameters. All parameter posteriors are well constrained within relatively narrow parameter range except for
the precessing frequency of the orbit $Q$. 
The posterior probability distribution of $Q$ suggests that the precession orbit cannot have a period shorter than log (Q)=2.5 and any period longer than log (Q) $>$ 3 gives equally good fit. To present the best model, like other parameters, we chose the median of MCMC sample for log (Q).
The best-fit parameters are determined by medians of each parameter's MCMC samples as shown by the middle
vertical lines in each 1D marginal posterior distribution in Fig. 4, and are listed in Table 2.

In the model fitting of the velocity curve, we used average broad-line velocity offset (black data points in Fig. 5a) for blue and red broad-line components.
Each error bar in black data points is a squared sum of fitting error of blue and red component, 
velocity resolution of each spectral data, and uncertainty of narrow line velocity measurement. 
The black solid line is the best-fit velocity curve, and the grey dashed line is $v_{\mbox{\tiny off}}$. 
We find that the model velocity curve fits surprisingly well to the data points. 
Fig. 5b shows the orbit of the rSMBH from best-fit model parameters (red dots represent epoch of observations).
Note that the semi-major axis length in the plot is normalized to 1
and does not show a projection effect (due to the inclination angle) along the line of sight. 
The best-fit model parameters suggest that the rSMBH orbit is slowly precessing in the clockwise direction with a precession period of $\sim$3200 yrs.
The period of the orbit is $\sim$29.2 yrs and its eccentricity $e$ is 0.94.

\section{Discussion}

We have modeled the broad line emission velocity offset data of Mrk 1018 using recoiled SMBH kinematics. 
Although the observed data does not cover the full range of orbital period more than twice to confirm 
a repeatability of the kinematics, it is very encouraging that the model fits very well to the data.
In this section, we discuss other aspects of the observation of Mrk 1018 and propose the oscillation 
of rSMBH as a possible origin of the variation of AGN activity.  
 
\subsection{Change of Extinction}
Extinction can be caused by material along the line of sight or by an increased dust covering factor (CF) due to decreased AGN activity (Stalevski et al. 2016). 
If the obscuration is partially responsible for the changing look in Mrk 1018, 
we expect to see variations in the measured extinction.
The extinction value of the broad line regions can be estimated from the Balmer decrement assuming an intrinsic H$\alpha$/H$\beta$ 
ratio of 3.1 (i.e. A$_v=7.2 \times$ log((H$\alpha$/H$\beta$)$_{obs}$) - 3.55, where the (H$\alpha$/H$\beta$)$_{obs}$ is 
observed H$\alpha$/H$\beta$ ratio). The estimated extinction values are listed in Table. 1 and annotated in Fig. 5b along the rSMBH orbit.
We do not find large values of extinction when the spectral type of the AGN is Type 1 (A$_v$=1.4 and A$_v$=0.7 at epoch 1984.08 and 2000.73, respectively, and no extinction at other epochs).
Our result is generally consistent with that of Husemann et al. (2016) who found no extinction between 2010 and 2016 from their X-ray observations.
However, we find large values of extinction when the AGN is in Type 1.9: A$_v$=4.3 and 3.0 at epoch 1979.70 and 2015.03, respectively.  
We argue that the increased dust CF is more plausible explanation to the observed extinction values for different 
epochs since the large extinction value measured in 1979.70 was not measured again when the rSMBH returns to the same location in the 
orbit around 2008, although the random chance of the line of sight being blocked by `moving' material is still possible.

\subsection{Photometric Variation}
The variability of AGN activity can also be traced by carefully monitoring the change of AGN luminosity in imaging data.
We analyzed multi-epoch SDSS g-band (from 2004 to 2008) and SWIFT B-band (from 2008 to 2016) imaging data from archive. 
The magnitude of AGN and host galaxy were measured by 2 dimensional morphology decompositions using GALFIT (Peng et al. 2002) with two component models: 
i.e., a PSF for the AGN and a Sersi\'c model for the host galaxy. 
For the SDSS images, we used coadded Stripe 82 images to derive the best-fit parameter of Sersi\'c index and effective radius,
and fixed them for modeling all images (n=3.0 \& $r_e$=12\arcsec). 
For the SWIFT images, we used the image observed in 2016 to derive the best-fit parameters of Sersi\'c index and effective radius since this image is observed when the AGN activity was minimum.
The derived parameters (n=3.1 \& $r_e$=13\arcsec) were fixed for modeling all SWIFT images. 
Fixing these parameters for the host galaxy is a reasonable approach to ensure a robust measurement of host galaxy luminosity which should not vary with time.

After measuring the AGN and host galaxy magnitudes, we have traced the variation of AGN activity using the AGN to host galaxy luminosity ratio ($L_{AGN}/L_{host}$) 
assuming the host galaxy luminosity does not change with time. 
We used this method since some of the SDSS images are observed under non-photometric conditions. 
This method also eliminates errors caused by zero-point magnitude determinations. 
The $L_{AGN}/L_{host}$ as a function of epoch is plotted in Fig. 6, where the triangles and filled circles represent SDSS and SWIFT data, respectively. 
We find that the AGN activity remains strong until around 2010 and starts to decrease rapidly thereafter. 
A similar trend is also found in Fig. 3 of McElroy et al. (2016).

\subsection{Spectral Variation}

As mentioned earlier, the spectral type of the Mrk 1018 was either Type 1.9 or Type 2 from 1974 to 1979 and changed to Type 1 around 1984.
We do not have spectra earlier than 1974, 
but the available data tell us that the spectral type of the Mrk 1018 has stayed Type 1.9/2 at least 6 years (1974 to 1980).
In Fig. 2, the broad H$\alpha$ emission line always exists in each epoch whereas the broad H$\beta$ emission line appears when the AGN activity becomes strong.
Thus, it could be possible to study the variation of AGN activity by tracing the change of broad H$\beta$ emission line flux.
However, we can not trace the absolute flux value of broad H$\beta$ emission line for all spectra since only SDSS and MUSE data are flux-calibrated.
Instead, we traced the change in broad H$\beta$ emission line with respect to [O III] line (i.e. H$\beta$/[O III] line ratio).
Here, we assumed that the flux density of [O III] line does not change during the period of 1974 to 2015 
simply because the change in BLR does not have sufficient time to propagate to the NLR located at a typical distance of a few hundred pc to a few kpc.

Fig. 7 shows the H$\beta$/[O III] line ratio as a function of time where we find some interesting properties.
The broad H$\beta$ emission line intensity increases and decreases rapidly at the start and at the end of the cycle, respectively:
it increased significantly from epoch 1979.70 to 1984.07 (about 10 times in $\sim$4 years),
and decreased significantly from epoch 2010.94 to 2015.03 (about 8 times in $\sim$4 years).
The broad H$\beta$ emission line intensity peaks around 2007.95, which is supported by steep continuum slope in blue part of the CTIO spectra (Fig. 2).
The redshifted H$\beta$ line falls in the middle of B and V filters, and 
Winkler (1997) observed that the B and V magnitudes of the Mrk 1018 dropped from Sep. 1992 to Dec. 1994.
This suggests that the H$\beta$ line may have two peaks: a small peak around 1984 (at the start of the cycle) and a large peak around 2008 (near the end of the cycle).
The two peaks could be explained by a perturbation in the accretion disk caused by the pericenteric passage of the rSMBH as we will discuss below.

\subsection{\lya\ Absorption}  
If the rSMBH is moving along its orbit, we may see spectroscopic evidence of obscuring material along the line of sight at random locations.
\lya\ line has been measured in 4 different epochs: 1984, 1986 by IUE, 1996 by HST/GHRS, 2016 by HST/COS.
From comparisons of \lya\ spectra observed in 1996 and 2016, Husemann et al. (2016)
detected an additional absorption line at $\lambda=1264\angstrom$ in 2016 spectra.
The epoch of 1996 is when rSMBH is near the apocenter and 
the epoch of 2016 is after the rSMBH passes the pericenter, but is still close to the center of host galaxy (see Fig. 5b).
This additional Lya absorption occurring in 2016 that was not seen in 1996 supports our rSMBH scenario 
that the rSMBH broad line region has higher chance of being blocked by line of sight material
in host galaxy when it is closer to the center, while it is less likely to be blocked when it is far away from the host galaxy center.
The obscuration material was blueshifted by $\sim 700\ \kms$ with respect to systemic velocity.
Husemann et al. (2016) suggested that the absorption was caused by neutral gas within the host galaxy since it has a short-time variability
and may not have caused the spectral changes in the Mrk 1018 because of its low column density ($N_H < 10^{19}$ cm$^{-2}$).

\subsection{Perturbation of Accretion Disk}
Based on all observed aspects of Mrk 1018 discussed above, 
we argue that an oscillating rSMBH is a very plausible scenario for the change in AGN activity in Mrk1018. 
Perturbations in the AGN accretion disk can be used to explain 
continuum flux changes and subsequent spectral changes (i.e. Stalevski, Jovanovic, \& Popovic 2008; Valtonen et al. 2009; Popovi\'c et al. 2014). 
The perturbation of the accretion disk induces disk instability, and amongst several mechanisms,
thermal instability is commonly invoked to explain AGN luminosity variations. 
The thermal instability time scale $\tau_{\mbox{th}}$ is given by (Liu et al. 2008)
\begin{equation}
\tau_{\mbox{th}} = 5 (\frac{\alpha}{0.1})^{-1} (\frac{M_{BH}}{10^8 M_{\odot}}) ({\frac{r_d}{10^3 r_g}})^{3/2} \mbox{yr},
\end{equation}   
where $r_d$ and $r_g$ are radius of accretion disk and gravitational radius ($=GM_{BH}/c^2$), respectively.
For standard $\alpha$-disk with $\alpha=0.1$ with typical disk radius of $r_d$=0.004 pc (Jim\'enez-Vicente et al. 2014) and black hole mass $M_{BH}$=10$^{8.15}M_{\odot}$ (Bennert et al. 2011), 
the thermal fluctuation lasts for only 4 years and thus it is difficult to explain the variation over a 35 year time scale unless an unusually small $\alpha$ value of 0.01 is used.

Another possible mechanism we can consider to explain the change in AGN activity is a tidal impulse that rSMBH experiences when it passes pericenter.
When rSMBH accretion disk passes pericenter, it feels tidal impulses from the mass of the host galaxy in the center since orbital velocity 
is much larger than the rotational velocity in the accretion disk.
The tidal impulse, which becomes stronger if the orbit has a larger eccentricity, 
causes a density perturbation in the accretion disk that will change the accretion rate and 
possibly the disk geometry. 
This perturbation can be erased by relaxation processes with a time scale given by sound crossing 
time $\tau_s$ (Liu et al. 2008)
 
\begin{equation}
\tau_{\mbox{s}} = 70 (\frac{M_{BH}}{10^8 M_{\odot}}) ({\frac{r_d}{10^3 r_g}}) (\frac{T}{10^5}K)^{-1/2} \mbox{yr}.
\end{equation} 
The effective temperature of accretion disk $T$ depends on the black hole mass and Eddington ratio, and can be estimated 
by
\begin{equation}
T=10^{5.56} (\frac{L_{bol}}{L_{Edd}})^{1/4} (\frac{M_{BH}}{10^8 M_{\odot}})^{-1/4} \mbox{K}
\end{equation}
for a rotating Kerr black hole (Bonning et al. 2007).
The effective temperature $T$ for Kerr black hole becomes $T=1.5\times10^5$K with $M_{BH}$=10$^{8.15}M_{\odot}$ (Bennert et al. 2011) and Eddington ratio ${L_{bol}}/{L_{Edd}}=0.03$ (McElroy et al. 2016).
For Schwarzschild black hole, $T$ will be lowered by factor of $10^{0.46}$ (Shields 1989) and becomes $T=0.5\times10^5$K.
If we use an average temperature of $T=10^5$K, $r_d$=0.004 pc, and $M_{BH}$=10$^{8.15}M_{\odot}$,
we can obtain a relaxation time (= sound crossing time) of $\sim$70 years.

As the rSMBH passes the pericenter, the accretion disk will be perturbed by a strong tidal force 
as similarly observed in an accretion disk in close binary system (Godon 1997; Makita et al. 2000). 
In contrast to the accretion disk in close binary system that subjects a constant tidal force from a companion star,
the accretion disk in rSMBH is subject to a tidal impulse since it rapidly passes the pericenter (i.e., host galaxy center) along the eccentric orbit.
Then the accretion rate that possibly has dropped after pericentric passage will start to increase as the perturbation is being erased over the relaxation time scale.
This time scale is reasonably long enough so that the accretion rate keeps increasing until 
the rSMBH passes the pericenter again where the accretion disk feels the tidal impulse and drops again.
This behavior can explain the trend of AGN activity in Mrk 1018; (1) increasing before and right after pericenter passage 
(1979-1986), (2) decreasing during 1992-1994 after pericenter passage, (3) increasing again until 2009-2010 and (4) decreasing after pericenter passage (2015-2017).

A recoiled SMBH experiencing a tidal impulse proposed in this work is a possible scenario that could explain the change of AGN activity in Mrk 1018. 
To confirm or rule out this hypothesis, we need more spectral and photometric data covering at least another full cycle.
However, if it is confirmed, it will have important implications: the relation between the observed broad-line velocity offsets and the change in AGN activity, accretion disk physics, and ultimately AGN unification theory. 
The AGN in post-merger systems could be either binary SMBHs or a rSMBH. 
For rSMBHs, broad line kinematics can be modeled as we have done in this work. 
For binary SMBH, broad-line kinematics for blue and red component can be modeled by a combination of circular
motion of two SMBHs and the motion of their center of mass which is not necessarily stationary since 
the binary SMBH can be still in the process of settling in the center of host galaxy.
Currently, there are a handful of AGNs with spectroscopic data spanning over 10 years. 
We are planning to model broad-line velocity kinematics of these systems.
We also plan to study the parameter space of rSMBH kinematics predicted from simulations 
with large samples of potential rSMBH candidates by long term monitoring in the LSST era. 
Detailed simulations of accretion disks affected by tidal impulse would be another interesting study to shed 
light on accretion physics of SMBH. 
   
\section{Summary}

Archival images and spectra were analyzed to investigate the physical mechanisms responsible for spectral change in Mrk 1018.
The following summarizes what we have found:

\noindent
\textbf{\emph{1) Properties of orbit:}}

$\bullet$
Two kinematically distinct blueshifted and redshifted broad-line components are found from spectral decomposition analysis.

$\bullet$
The velocity offset curve of the broad-line components as a function time shows characteristic pattern
and can be explained by oscillating rSMBH.

$\bullet$
Our Bayesian MCMC simulation suggests the rSMBH has a highly eccentric orbit ($e$=0.94) with orbital period of 29.2 years.

\noindent
\textbf{\emph{2) Nature of AGN variations:}}

$\bullet$
Monitoring of the intensity variation in nuclear region suggests that the AGN activity started to decrease rapidly from $\sim$2010.

$\bullet$
Monitoring of the flux variation of broad H$\beta$ emission line suggests that the AGN activity rises and drops rapidly at the start and at end of the cycle. 
It also suggests an existence of two activity peaks: a small peak at the start and a large peak near the end of the cycle.

$\bullet$
The extinction value was found to be increased at Type 1.9 stage probably due to an increased covering factor.

$\bullet$
Spectral data suggest that the period of a full cycle from Type 1.9/2 to Type 1 and returning Type 1.9 may be at least 40 years (from 1975 to 2015).

We found that the nature of AGN variations including spectral change can be explained by 
a tidal impulse the accretion disk receives when the rSMBH passes the pericenter.
If our explanation for the spectral change in AGN is confirmed by subsequent observations, this phenomenon would have a clear impact on our understanding of AGN variability on short timescales.
We do not know if it will repeat a similar pattern of spectral change in the future,
but if it does, its spectral type could turn to Type 1 around mid 2020's.

The authors thank the anonymous refree for comments and suggestions that greatly improved this paper.
We also thank M. Crenshaw for generously sharing their Lowell and CTIO spectra, L. Blecha for reading the manuscript and providing feedback, M. Kim for flux calibrating the MUSE spectra, and C. Peng for helping with the Galfit.
D.C.K., I.Y. \& A.S.E. acknowledge support from the National Radio Astronomy Observatory (NRAO).
The National Radio Astronomy Observatory is a facility of the National Science Foundation operated under cooperative agreement by Associated Universities, Inc.

%
%

\clearpage

\begin{deluxetable}{lccccc}
\tabletypesize{\small}
\tablewidth{0pt}
\tablecaption{Results of spectral decomposition}
\tablehead{
\multicolumn{1}{c}{Source} &
\multicolumn{1}{c}{Epoch} &
\multicolumn{2}{c}{Velocity offset (FWHM) [$\kms$]} &
\multicolumn{1}{c}{$\rm{A_v}$} &
\multicolumn{1}{c}{Reference$^{*}$} \\
\multicolumn{1}{c}{} &
\multicolumn{1}{c}{} &
\multicolumn{1}{c}{Blue component} &
\multicolumn{1}{c}{Red component} &
\multicolumn{1}{c}{mag} &
\multicolumn{1}{c}{}
}
\startdata
Lick       & 1979.70 & -1190$\pm$40  (8120) &  1370$\pm$20  (2890) &  4.3   & 1      \\
UCSD/UMinn & 1984.08 & -2760$\pm$390 (5470) &    80$\pm$30  (3220) &  1.4   & 2      \\
HST        & 1996.77 &  -560$\pm$80  (2210) &  1570$\pm$60  (1900) &\nodata & 3      \\
SDSS       & 2000.73 &  -650$\pm$110 (2900) &  1370$\pm$70  (3430) &  0.7   & 4      \\
6df        & 2004.62 &  -770$\pm$50  (2640) &  1450$\pm$80  (4560) &  0.0   & 5      \\
CTIO       & 2007.77 & -1050$\pm$30  (1960) &  1400$\pm$130 (4530) &  0.0   & 6      \\
CTIO       & 2007.95 & -1030$\pm$40  (1990) &  1250$\pm$140 (4590) &  0.0   & 6      \\
Lowell     & 2009.96 & -1050$\pm$120 (2020) &  1010$\pm$300 (2630) &  0.0   & 6      \\
Lowell     & 2010.94 & -1160$\pm$50  (2370) &   730$\pm$210 (2490) &  0.0   & 6      \\
VLT/MUSE   & 2015.03 &  -650$\pm$40  (3110) &  3350$\pm$110 (4370) &  3.0   & 7      \\
\enddata
\tablenotetext{*}{Reference: 1) Osterbrock 1981, 2) Cohen et al. 1986, 3) HST Proposal 6704 (PI: R. Goodrich), 4) Abazajian et al. 2009, 5) Jones et al. 2009, 6) Trippe et al. 2010, 7) McElroy et al. 2016.}
\end{deluxetable}

\begin{deluxetable}{lcccccc}
\tabletypesize{\small}
\tablewidth{0pt}
\tablecaption{Best fit rSMBH orbital parameters}
\tablehead{
\multicolumn{1}{c}{$P$ (yr)} &
\multicolumn{1}{c}{$e$} &
\multicolumn{1}{c}{$t_0$ (yr)} &
\multicolumn{1}{c}{$a$ (pc)} &
\multicolumn{1}{c}{$Q$ (yr)} &
\multicolumn{1}{c}{$w_0$ (rad)} &
\multicolumn{1}{c}{$v_{off}$ (\kms)} 
}
\startdata
29.2  & 0.94 & 4.85  & 0.004 & 3200 & -0.31 & 460   \\
\enddata
\end{deluxetable}

\clearpage

\begin{figure}
\centerline{\includegraphics[scale=0.7]{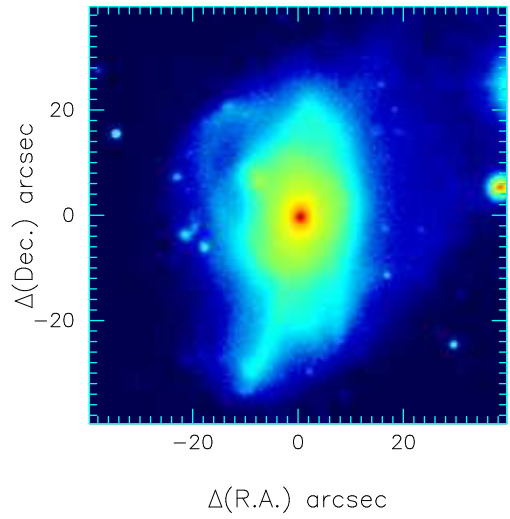}}
\caption{SDSS Stripe 82 image of Mrk 1018. The image shows signatures (a tidal tail and disturbed envelope) of merger event. 
In this figure, North is top and East is to the left.
}
\end{figure}

\begin{figure}
\centerline{\includegraphics[scale=1]{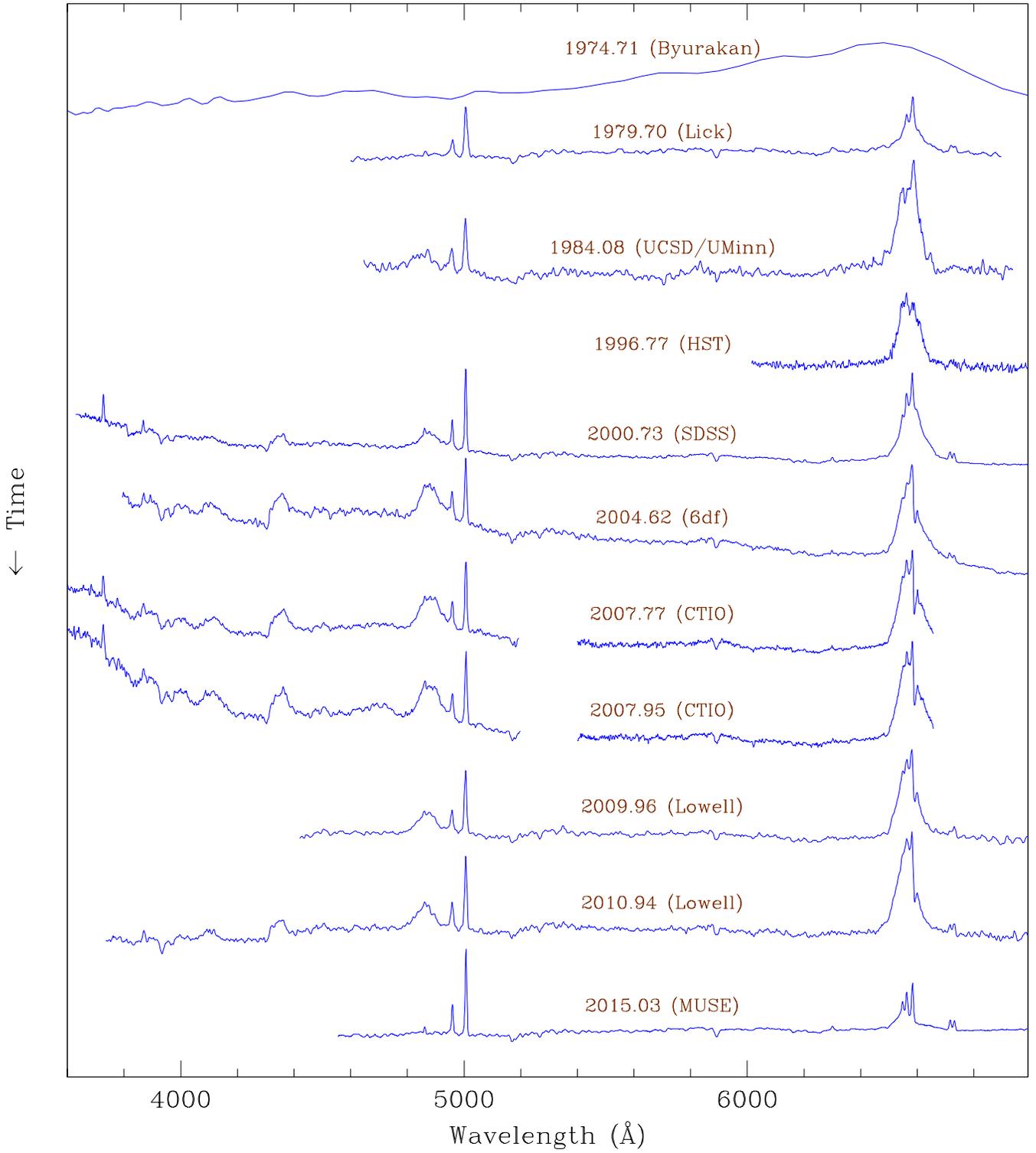}}
\caption{Archival spectra of Mrk 1018 used for spectral analysis. All spectra, except for the Byurakan spectra, are normalized to have the same integrated flux value of [O III]$\lambda 5007$\AA \ line.
}
\end{figure}

\begin{figure}[!ht]
\centerline{\includegraphics[scale=0.7]{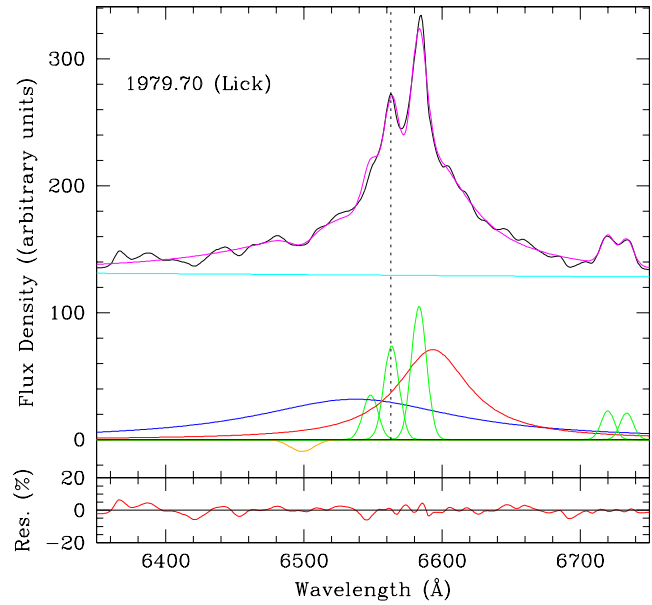}}
\caption{Result of spectral decompositions for H${\alpha}$ $+$ [N II] lines.
Black, cyan, blue, red, green, orange, and magenta lines represent
data, power-law continuum, broad emission line blue component, broad emission line red component, narrow emission lines, absorption line, and model (sum of all fitting components), respectively.
Vertical dotted line in the plot represents line center of narrow H${\alpha}$ emission line.
The residual of the fitting (data/model in percentage) is shown on the bottom of the plot.
An extended version of this figure is available online.
}
\end{figure}

\clearpage
\begin{figure}
\centering
\epsfig{figure=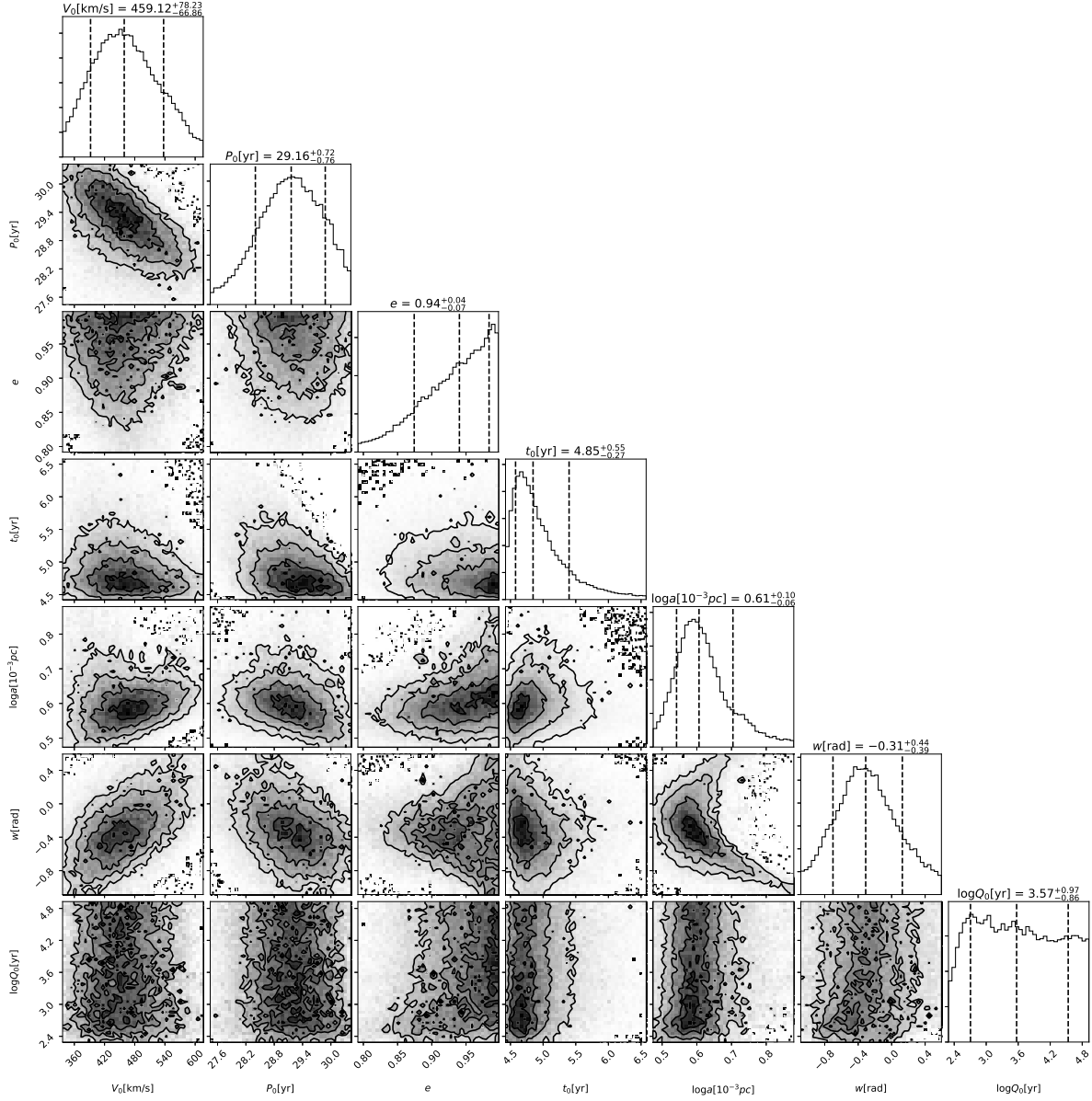,scale=0.4}
\caption{Posterior probability distribution of model parameters. Diagonal plots show 1D marginal posterior distribution of each model parameter 
with three vertical lines indicating 16, 50 and 84\% quantile respectively, and off-diagonal plots show 2D joint marginal posterior distribution of 
two model parameters.
}\label{fig:param_fit}
\end{figure}

\clearpage
\begin{figure}
\centering
\epsfig{figure=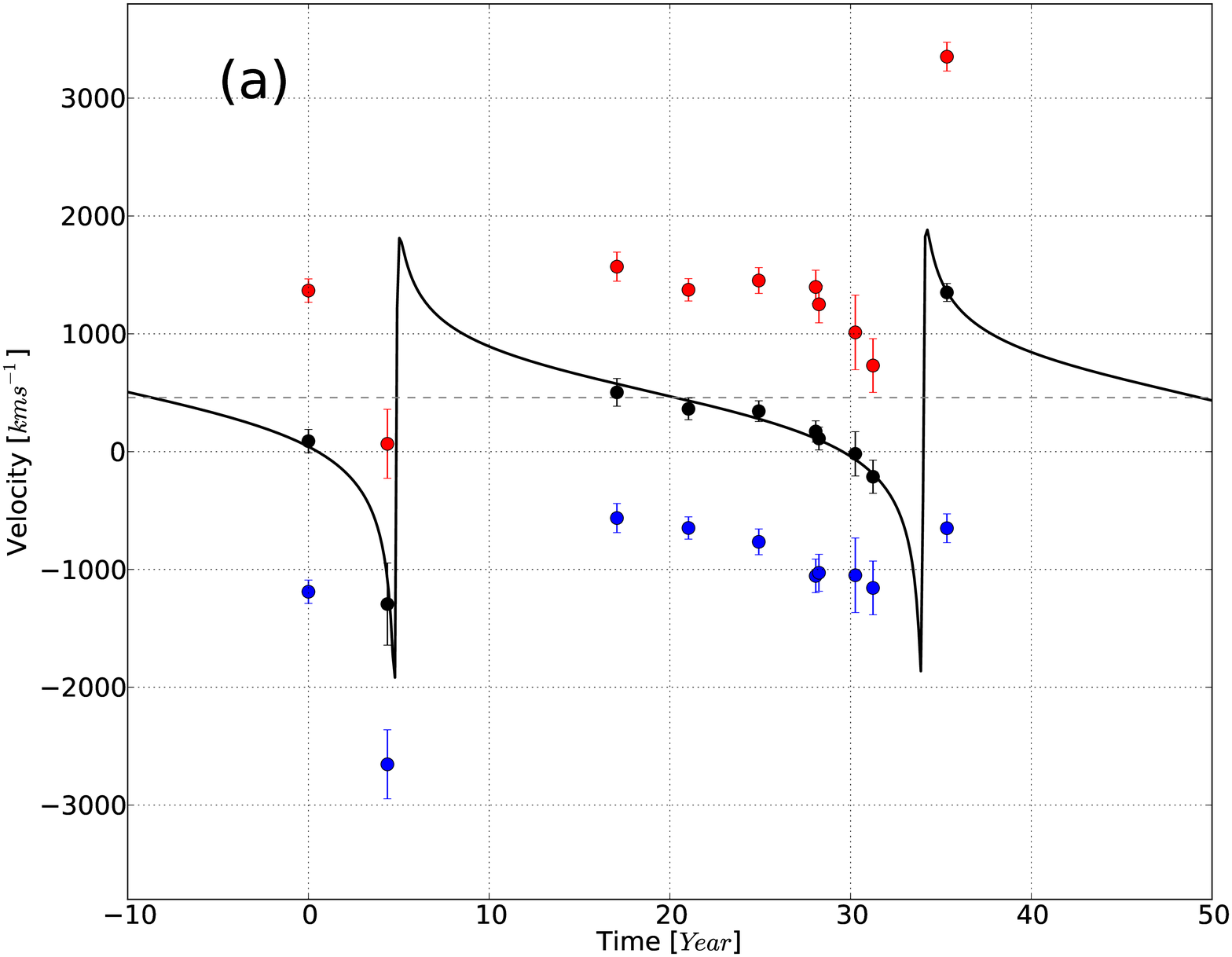,scale=0.4}\vspace{-0.5cm}
\epsfig{figure=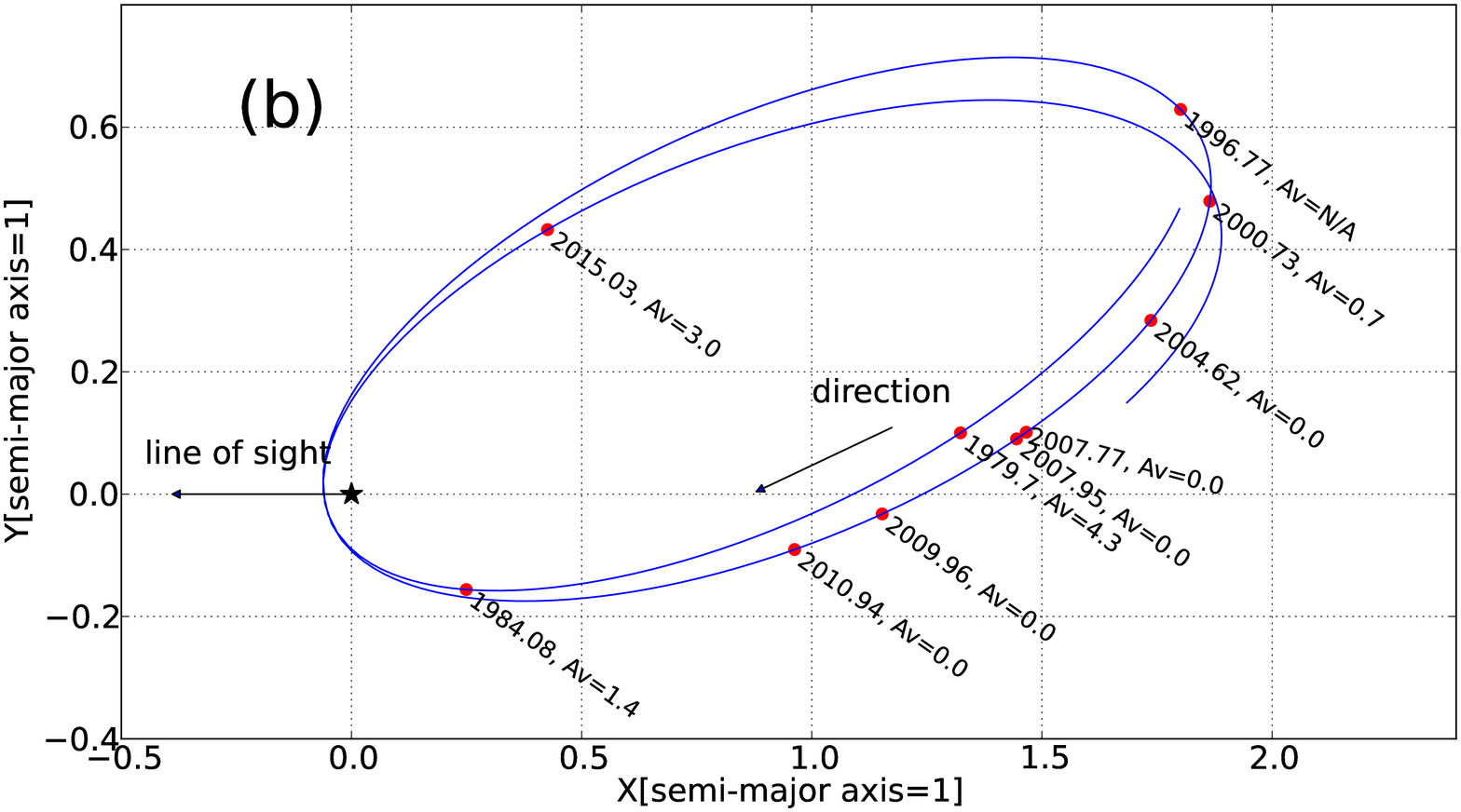,scale=0.4}
\caption{Panel(a): Broad-line velocity offset as a function epoch, where blue, red, and black represent velocity offset for blue component, red component, and average of blue and red components, respectively. 
The black line shows best-fit velocity curve and the grey dashed line indicates the best fit value of $v_{off}$ in Eq. (4).
Panel(b): Orbit of rSMBH from best-fit model parameters where each epoch is marked along the orbit.
The X and Y coordinates are normalized with respect to semi-major and semi-minor axis, respectively.
}\label{fig:pos_and_vel}
\end{figure}

\begin{figure}[!ht]
\centerline{\includegraphics[scale=1.0]{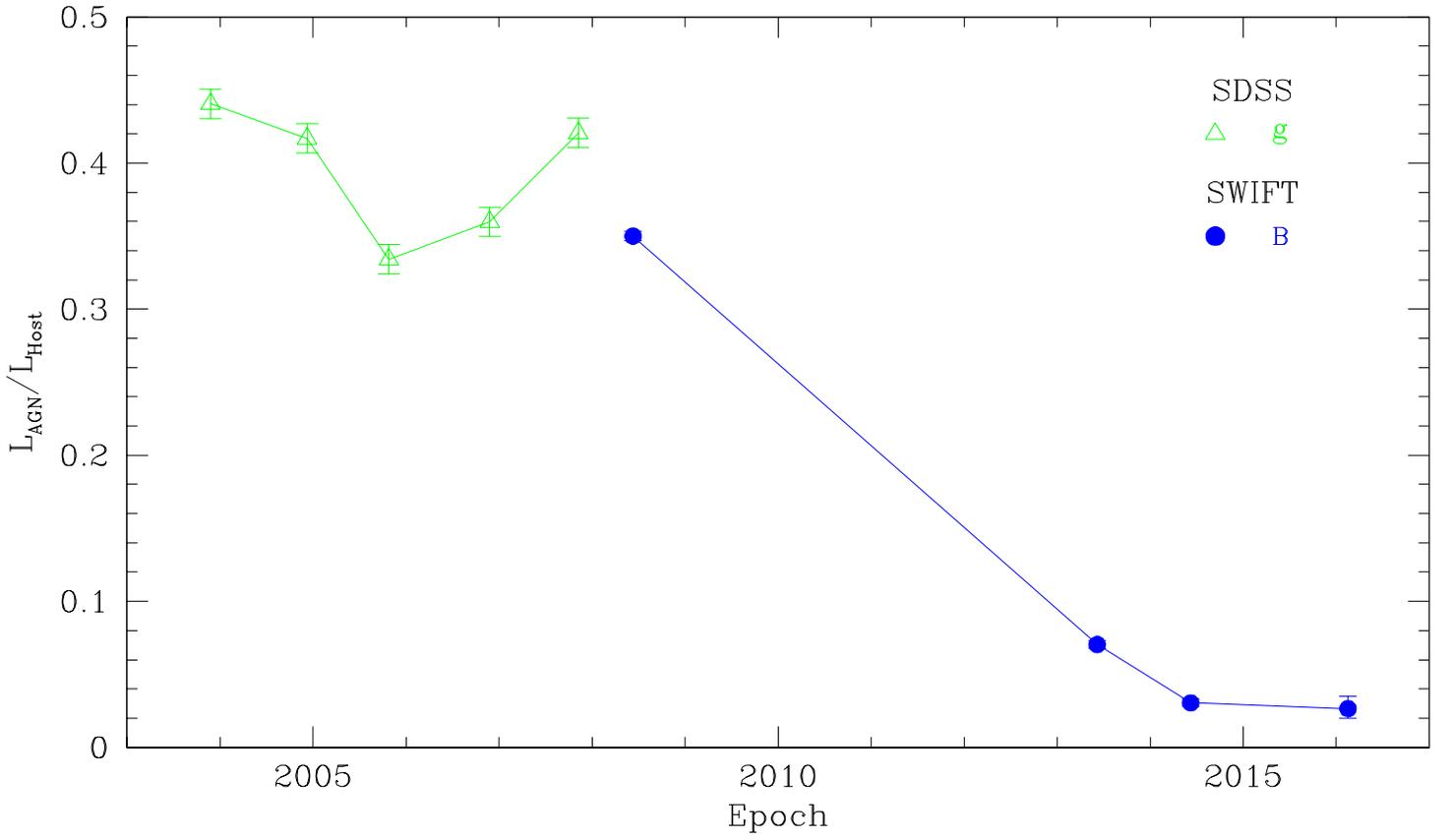}}
\caption{Luminosity ratio of AGN (L$_{\rm{AGN}}$) to host galaxy (L$_{\rm{Host}}$) as a function epoch.
Open triangles and filled circles represent SDSS and SWIFT data, respectively.
}
\end{figure}

\begin{figure}[!ht]
\centerline{\includegraphics[scale=0.9]{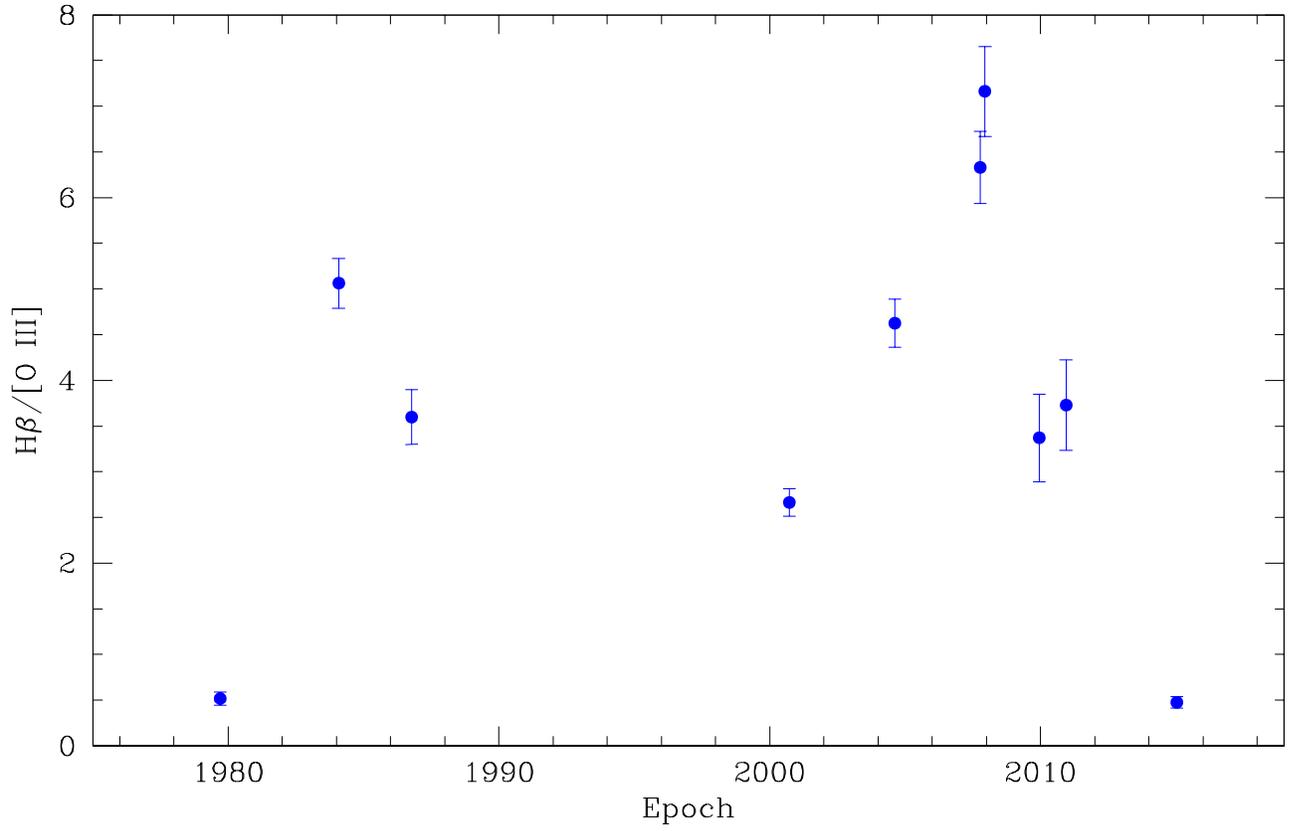}}
\caption{Variation of \hb\ broad-line intensity normalized to [O III] line as a function of epoch.
Data point at epoch 1986.78 is from Salzer et al. (1989).
}
\end{figure}

\end{document}